\documentclass[12pt]{article}
\newcommand{\br}{{\bf R}}
\newcommand{\ls}{{\cal L}_{{\rm s}}}

\newcommand{\Tr}{\rm{Tr}}

\usepackage{latexsym,amsfonts,amsmath,amsthm,amssymb}

\usepackage[T1]{fontenc}
\usepackage[latin1]{inputenc}
\title{Quantum mechanics as an asymptotic projection of statistical mechanics of classical fields}
\author{Andrei Khrennikov\\
International Center for Mathematical
Modeling \\ in Physics and Cognitive Sciences,\\
University of V\"axj\"o, S-35195, Sweden}

\begin{document}
\maketitle

\abstract{We show that quantum mechanics can be represented as an
asymptotic projection of statistical mechanics of classical fields.
Thus our approach does not contradict to a rather common opinion
that quantum mechanics could not be reduced to statistical mechanics
of classical particles. Notions of a system and causality can be
reestablished on the prequantum level, but the price is sufficiently
high -- the infinite dimension of the phase space. In our approach
quantum observables, symmetric operators in the Hilbert space, are
obtained as derivatives of the second order of functionals of
classical fields. Statistical states are given by Gaussian ensembles
of classical fields with zero mean value (so these are vacuum
fluctuations) and dispersion $\alpha$ which plays the role of a
small parameter of the model (so these are small vacuum
fluctuations). Our approach might be called {\it Prequantum
Classical Statistical Field Theory} - PCSFT. Our model is well
established on the mathematical level. However, to obtain concrete
experimental predictions -- deviations of real experimental averages
from averages given by the von Neumann trace formula - we should
find the energy scale $\alpha$ of prequantum classical fields.}

\section{Introduction}
Since the first days of creation of quantum mechanics, physicists, mathematicians and philosophers
are involved in stormy debates on the possibility to create a classical prequantum statistical model,
see for example [1]--[44]. Here ``classical statistical'' has the meaning of a realistic model
in that physical variables can be considered as objective properties and probabilities can be described by
 the classical  (Kolmogorov) measure-theoretic model.
 There is a rather common opinion that it is impossible to construct such a prequantum model.
Such an opinion is a consequence of Bohr's belief that quantum
mechanics is a {\it complete theory.} By the orthodox Copenhagen
interpretation it is in principle impossible to create a deeper
description of physical reality. In particular, there is a rather
common belief that quantum randomness is irreducible, see e.g. von
Neumann [4] (in the opposite to classical randomness which is
reducible in the sense that it can be reduced to ensemble randomness
of objective properties). There is a huge activity in proving
various mathematical "NO-GO" theorems (e.g. von
Neumann\footnote{Recently A. Leggett paid my attention to the fact
that J. von Neumann did not consider his derivations [4] as a
rigorous mathematical theorem. In the original German addition
(1933) of his book [4] he called "NO-GO" considerations ``ansatz''
and not theorem.}, Kochen-Specker, Bell, see, for example, [12],
[14], [15], [31] for details). Many people think that with the aid
of such mathematical investigations it is possible to prove
completeness of quantum mechanics. As I pointed out in the preface
to the conference proceedings [36], such an approach can not be
justified, because we do not know {\it correspondence rules} between
prequantum and quantum models. Therefore each attempt to formulate a
new ``NO-GO'' theorem is in fact an attempt to present a list of
properties of a classical $\to$ quantum map $T.$

First time such a list was presented by  J. von Neumann  in his book
[4]. Later his list was strongly criticized by many authors
(including J. Bell [12], also L. Ballentine [15]). In particular,
there was criticized the assumption on {\it on-to-one
correspondence} between the set of classical physical variables $V$
and the set of quantum observables $O.$ There was also pointed out
that von Neumann's assumption that
$$T(a+b)= T(a)+ T(b)$$
 for any two
physical variables (so without the assumption that observables
$T(a)$ and $T(b)$ can be measured simultaneously) is nonphysical.
Then different authors proposed their own lists of possible features
of the map $T$ which (as they think) are natural. These lists
(including Bell's list) were again criticized, see e.g. [26],
[27]--[29], [37], [40], [41], [44] and some papers in [33]--[36].

In [45] I proposed to start the activity in the opposite direction.
Instead of looking for lists of assumptions on the classical $\to$
quantum map $T$ which would imply a new ``NO-GO'' theorem, it seems
to be more natural to try to find such lists of features of $T$
which would give the possibility to create a natural prequantum
classical statistical model. In these papers it was shown that all
distinguishing features of the quantum probabilistic model
(interference of probabilities, Born's rule, complex probabilistic
amplitudes, Hilbert state space, representation of observables by
operators) are present in a latent form in the classical Kolmogorov
probability model. The main problem was that the construction of
quantum representation of the classical probability model [45] was
purely mathematical (probabilistic). The main task was to find a
concrete natural (from the physical point of view) classical
statistical model $M$ which would reproduce QM. We present such a
prequantum model in this paper. Our model is classical statistical
mechanics on the Hilbert phase-space. Points of this phase-space can
be considered as {\it classical fields} (if we take the Hilbert
space $H=L_2({\bf R}^3)).$ Our approach might be called {\it
Prequantum Classical Statistical Field Theory} - PCSFT.

Our approach is an asymptotic approach. We introduce a small
parameter $\alpha$ -- dispersion of ``vacuum fluctuations''; so in
fact we consider a one parameter family of classical statistical
models $M^\alpha.$ QM is obtained as the limit of classical
statistical models when $\alpha \to 0:$
\begin{equation} \label{CP}
 \lim_{\alpha\to 0}M^\alpha=N_{\rm{quant}},
\end{equation}
where $N_{\rm{quant}}$ is the Dirac-von Neumann quantum model [2],
[4]. We pay attention that our approach should not be mixed with so
called {\it deformation quantization}, see e.g. [47]--[49]. In the
formalism of deformation quantization classical mechanics on the
phase-space $\Omega_{2n} = {\bf R}^{2n}$ is obtained as the
$\lim_{h\to 0}$ of quantum mechanics (the correspondence principle).
In the deformation quantization the quantum model is considered as
depending on a small parameter $h:N_{\rm{quant}}\equiv
N_{\rm{quant}}^h,$ and formally
\begin{equation}
\label{CP1}
 \lim_{h \to 0} N_{\rm{quant}}^h = M_{\rm{conv. class.}}
\end{equation}
where  $M_{\rm{conv. class.}}$ is the conventional classical model
with the phase-space $\Omega_{2n}.$ In the opposition to deformation
quantization, we study the inverse problem, namely classical $\to$
quantum correspondence.

The crucial point is that our prequantum classical  statistical
model is not  conventional classical statistical mechanics on the
phase-space $\Omega_{2n}= {\bf R}^{2n},$ but its
infinite-dimensional analogue. Here the phase-space $\Omega= H
\times H,$ where $H$ is the (real separable) Hilbert space. Thus
{\it the price of realism and causality is the infinite-dimension of
the phase-space!}

In our approach the classical $\to $ quantum correspondence $T$ is
based on the Taylor expansion of classical physical variables --
functions $f: \Omega \to {\bf R}.$ This is a very simple map:
function is mapped into its second derivative (which is always a
symmetric operator).\footnote{By the terminology which is used in
functional analysis $f$ is called functional -- a map from a
functional space into real numbers. If we represent $\Omega$ as the
space of classical fields, $\psi:{\bf R}^3 \to {\bf R},$ then
$f(\psi)$ is a functional of classical field.}

The space of classical statistical states consists of Gaussian
measures on $\Omega$ having zero mean value and dispersion $\approx
\alpha.$ Thus a statistical state $\rho$ (even a so called pure
state $\psi\in \Omega, \Vert \psi \Vert=1)$ can be interpreted as a
Gaussian ensemble of classical fields which are very narrow
concentrated near the vacuum field, $\psi_{\rm{vacuum}}(x)=0$ for
all $x \in {\bf R}^3.$ Such a $\rho$ has the very small standard
quadratic deviation from the field of vacuum $\psi_{\rm{vacuum}}:$
\begin{equation}
\label{SD} \int_{L_2({\bf R}^3 \times L_2({\bf R}^3)} \int_{{\bf
R}^3} [p^2(x) + q^2(x)] dx d\rho (q,p) = \alpha, \; \alpha\to 0,
\end{equation}
where a classical (prequantum) field $\psi(x)$ is a vector field
with two components $ \psi(x)=(q(x), p(x)).$ Suppose that square of
the field has the dimension of energy (as in the case of
electromagnetic field in the Gaussian system of units). Then a
statistical state $\rho$ is en ensemble of fluctuations of vacuum
which are small in the energy domain.

The choice of the space of statistical states plays the crucial role
in our approach. QM is the image of a very special class of
classical statistical states. Therefore we discuss this problem in
more detail. Let us use the language of probability theory. Here a
statistical state is represented by a Gaussian random variable $
\lambda \to \psi_\lambda,$ where $\lambda$ is a random parameter. We
have:
\begin{equation}
\label{SD1} E \psi_\lambda=0, \sigma^2(\psi)=  E \vert \psi_\lambda
- \psi_{\rm{vacuum}} \vert^2  =\alpha.
\end{equation}
We pay attention to the evident fact that small dispersion does not
imply that the random variable $\psi(\lambda)$ is small at any point
$\lambda \in \Lambda.$  The total energy of the field $\psi$
$$
{\cal E} (\psi_\lambda)\equiv \int_{{\bf R}^3} \vert \psi_\lambda(x)
\vert^2 d x = \int_{{\bf R}^3} [p^2_\lambda (x) + q^2_\lambda(x)] dx
$$
can be arbitrary large (with nonzero probability). But the
probability that ${\cal E} (\psi_\lambda)$ is sufficiently large is
very small. The easiest way to estimate this probability is to use
the (well known in elementary probability theory) Chebyshov
inequality:
\begin{equation}
\label{SD2} P( \lambda : {\cal E} (\psi_\lambda) > C) \leq E {\cal
E}(\psi_\lambda)/C= \alpha/C \to 0, \alpha\to 0,
\end{equation}
for any constant $C>0.$

It is important to remark that energy of the prequatum field $\psi$
should not be identified with energy of a quantum particle, e.g., an
electron or photon. The ${\cal E}(\psi)$ is the {\it internal
energy} of the prequantum field $\psi.$ Quantum particles appear in
our model as the result of interaction of the $\psi$-field and
various potentials.

It is especially interesting  that in our approach ``pure quantum
states'' are not pure at all! These are also statistical mixtures of
small Gaussian fluctuations of the ``vacuum field''. It seems that
the commonly supported postulate, see e.g.  von Neumann [4], about
{\it irreducible quantum randomness}, i.e. randomness which could
not be reduced to classical ensemble randomness, was not justified.

At the moment we are not able to estimate the magnitude $\alpha$ of
Gaussian vacuum fluctuations. In the first version of our work [50]
we assumed, as it is common in SED [51], [52] as well as in Nelson's
stochastic QM [18], that $\alpha$ has the magnitude of  the Planck
constant $h.$ However, we could not justify this fundamental
assumption on the magnitude of vacuum fluctuations in our approach.
We could not exclude the possibility that
\begin{equation}
\label{CPP} \alpha/h <<1
\end{equation}
In such a case our approach  to vacuum fluctuations essentially
deviates from SED and Nelson's stochastic QM. One might even
speculate on a connection with {\it cosmology and string theory.}
However, in the present paper we consider the magnitude of vacuum
fluctuations just as  a small parameter of the model: $\alpha \to
0.$

We shall discuss a possible physical interpretation of our model and
its relation to other realistic prequantum models in section 9:  the
pilot wave model (Bohmian mechanics), see e.g. [11], [7], [19], SED
and Nelson's stochastic QM, see e.g. [18], [51], [52] and references
thereby.  In our model prequantum reality is reality of classical
fields - systems with the infinite number of degrees of freedom. So
we consider a field model, but this is not QFT, because quantum
mechanics is reproduced not through quantum, but classical fields.

This paper also can be considered as a contribution to the old
debate on {\it incompleteness of quantum mechanics} that was started
more than 70 years ego by Einstein, Podolsky and Rosen, see their
famous paper [5] on the ``EPR-paradox''. Our investigation supports
the EPR-thesis on incompleteness of quantum mechanics. In our model
both the position and momentum operators, $\hat{q}$ and $\hat{p},$
represent the ''elements of physical reality'':  not reality of
particles, but reality of fields. In PCSFT the $\hat{q}$ and
$\hat{p}$ are images of functionals of classical fields, $f_q(\psi)$
and $f_p(\psi).$ Moreover, in our approach quantum mechanics is not
complete, because our ontic model (describing reality as it is)
contains even statistical states having dispersions $\sigma^2
(\rho)= o(\alpha).$ Such statistical states are neglected in the
process of classical $\to$ quantum correspondence. QM does not
contain images of these states (as it was pointed out QM contains
only images given by density matrices of Gaussian states having the
dispersions of the magnitude $\alpha).$

Our approach is very close to attempts of E. Schr\"odinger [6], [7]
and A. Einstein [8], [9] to create purely (classical) field model
inducing quantum mechanics.

The main experimental prediction of PCSFT is that QM does give only
approximations of statistical averages. Therefore in principle there
could be found deviations of experimental averages from averages
predicted by QM and calculated through the von Neumann trace-formula
for averages. To get the concrete prediction, we should know the
magnitude of the prequantum field fluctuations $\alpha$ (which is
just a small parameter in our theoretical paper).

\section{Classical $\to $ quantum correspondence}

\subsection{Ontic and epistemic models} We show that (in the opposition to the very
common opinion) it is possible to construct a prequantum classical
statistical model. From the very beginning we should understand that
prequantum and quantum models give us two different levels of
description of physical reality. We can say that prequantum and
quantum models provide, respectively, {\it ontic} and {\it
epistemic} descriptions. The first describes nature as it is (as it
is ``when nobody looks''). The second is an observational model. It
gives an image of nature through a special collection of
observables, cf. [6], [7], [37], [53 -- 56]. QM is an example of an
epistemic model of nature. In fact, this was the point of view of N.
Bohr and W. Heisenberg and many other adherents of the Copenhagen
interpretation, see [56] for an extended discussion. The only
problem for us is that the majority of scientists supporting the
Copenhagen interpretation deny the possibility to create a
prequantum ontic model which would reproduce (in some way) quantum
averages. We recall that   it was not the whole Copenhagen school
that ``denied'' the possibility to create prequantum
models.\footnote{Recently Andrei Grib told me that at the beginning
Nils Bohr was not so much excited by the system of views that was
late known as the Copenhagen interpretation. Only dialogs with
Vladimir Fock convinced him in usefulness of such an interpretation.
Thus the Copenhagen interpretation is, in fact, the ``Leningrad
interpretation.'' Personal views of N. Bohr were presented in rather
unclear form, see [55], [56] for details.} Pauli, for example, just
believed that such approaches would take away the efficiency of
quantum formalism. The complete denial of these possibilities came
later, mostly under the influence of the theorem of Bell. In our
approach the ontic description is given by a continuous field-model,
cf. E. Schr\"odinger [6]:

{\small ``We do give a complete description, continuous in space and
time, without leaving any gaps, confirming the classical ideal of a
description of something. But we do not claim that this something is
the observed and observable facts.''}

In any ontic (``realistic'')  model there are given  the following
sets: a) $\Omega$ -- states; b) $V(\Omega)$ - physical variables.
Elements of $V(\Omega)$ desribe objective properties. In general it
is not assumed that they can be  measured. In a statistical ontic
model there are also considered {\it statistical states};  these are
distributions of states. Thus an {\it ontic statistical model} is a
pair $M= ( S(\Omega),  V(\Omega)),$ where $S( \Omega)$ a set of
statistical states of a model.

In any epistemic (``observational'') statistical model there are
given  sets of observables  and statistical states: $O$ and $D.$ An
{\it epistemic (``observational'')  statistical model} is a pair $N=
(D, O).$ Elements of  the set $O$ do not describe objective
properties; they describe results of observations. Statistical
states represent distributions of states $\omega \in \Omega.$ In
general (``individual states'') $\omega$ do not belong to the domain
of an epistemic model $N= (D, O)$ (because  observers using this
model in general are not able to prepare ``individual states''
$\omega).$ The set of states $D$ of $N$ need not contain images of
$\delta_\omega$-measures concentrated at points $\omega \in \Omega.$

Of course, in physics there are used some epistemic statistical models which describe even ``individual states'' (belonging to the domain of the corresponding ontic model). Here all measures $\delta_\omega, \omega \in \Omega,$ belong to the set of statistical states $D.$  However, in such a case one need not distinguish ontic and
epistemic levels of description.

For example, we can consider {\it classical statistical mechanics.}
Here states are given by points $\omega=(q,p)$ of the phase-space
$\Omega_{2n}= {\bf R}^{2n}$ and statistical states by probability
distributions on $\Omega_{2n}.$ States $\omega \in \Omega_{2n}$ can
be represented by statistical states --  $\delta_\omega$-measures on
the phase-space.

In the present paper we are not interested in such statistical
models. We are interested in epistemic models which do not provide a
description of ``individual states.'' In such a case
$\delta_\omega$-measures are not represented by statistical states
of an epistemic model: $D$ does not contain $T(\delta_\omega),$
where $T$ is a map performing correspondence between the ontic
(preobservational) model $M$ and the epistemic (observational) model
$N.$

We now discuss mathematical representations of ontic and epistemic models.

\subsection{Classical and quantum  statistical models}
Of course, there are many ways to
proceed mathematically both on the ontic and epistemic levels of
description of nature. But traditionally ontic models are
represented as {\ ``classical statistical models'':}

a).  Physical states $\omega$ are represented by points of some set $\Omega$ (state space).

b). Physical variables are represented by functions $f: \Omega \to
{\bf R}$ belonging to some functional space
$V(\Omega).$\footnote{The choice of a concrete functional space
$V(\Omega)$ depends on various physical and mathematical factors.}

c). Statistical states are represented by probability measures on
$\Omega$ belonging to some class $S(\Omega).$ \footnote{It is
assumed that there is given a fixed $\sigma$-field of subsets of
$\Omega$ denoted by $F.$ Probabilities are defined on $F,$ see A. N.
Kolmogorov [57], 1933. It is, of course, assumed that physical
variables are represented by random variables -- measurable
functions. The choice of a concrete space of probability measures
$S(\Omega)$ depends on various physical and mathematical factors.}

d). The average of a physical variable (which is represented by a function $f \in V(\Omega))$ with respect to a statistical state (which is
represented by a probability measure  $\rho \in S(\Omega))$ is given by
\begin{equation}
\label{AV0} < f >_\rho \equiv \int_\Omega f(\omega) d \rho(\omega) .
\end{equation}

\medskip

A {\it classical statistical model} is a pair $M=(S(\Omega),
V(\Omega)).$

\medskip

We recall that classical statistical mechanics on the phase space
$\Omega_{2n}$ gives an example of a classical statistical model. But
we shall not be interested in this example in our further
considerations. We shall develop  a classical statistical model with
{\it an infinite-dimensional phase-space.}

{\bf Remark 2.1.} {\small We emphasize that the space of variables $V(\Omega)$ need not coincide with the space of all random variables $RV(\Omega)$ -- measurable functions $\xi: \Omega \to {\bf R}.$ For example,
if $\Omega$ is a differentiable manifold, it is natural to choose $V(\Omega)$ consisting of smooth functions;
if $\Omega$ is an analytic manifold, it is natural to choose $V(\Omega)$ consisting of analytic functions and so on.
Denote the space of all probability measures on the $\sigma$-field $\Sigma$ by the symbol $PM(\Omega).$
The space of statistical states $S(\Omega)$ need not coincide with $PM(\Omega).$ For example, for some statistical
model $S(\Omega)$ may consist of Gaussian measures.}

We shall be interested in ontic models (which are mathematically represented as classical statistical models) inducing  the quantum epistemic (observational) statistical model $N_{\rm{quant}}.$

In the Dirac-von Neumann formalism [2], [4] in the complex Hilbert
space $H_c$ this model is described in the following way:

\medskip

a). Physical observables are represented by operators $A: H_c \to H_c$ belonging to the
class of continuous\footnote{To simplify considerations, we shall  consider only quantum observables
represented by bounded operators. To obtain the general quantum model with
observables represented by unbounded operators, we should consider a prequantum classical
statistical model based on the Gelfand triple: $H_c^{+} \subset H_c \subset H_c^{-}.$} self-adjoint operators $\ls \equiv \ls (H_c)$ (so $O$ is mathematically represented by $\ls).$

b). Statistical states are represented by density operators,  see [4]. The class of such operators
is denoted by  ${\cal D} \equiv {\cal D} (H_c)$ (so $D$ is mathematically represented by ${\cal D}).$

d). The average of a physical observable (which is represented by the operator $A \in \ls (H_c))$ with respect to a statistical state (which is represented  by the density operator $D \in {\cal D} (H_c))$ is given by von Neumann's
formula:
\begin{equation}
\label{AV1}
<A >_D \equiv \Tr\; DA
\end{equation}

\medskip

The {\it quantum statistical model} is the pair $N_{\rm{quant}}
=({\cal D}(H_c), \ls(H_c)).$

\medskip

Typically in quantum formalism there are also considered  {\it pure
quantum states} given by normalized vectors $\psi \in H_c,$ see [2],
[4]. We shall not do this, because ``pure states'' of conventional
quantum mechanics do not coincide with ontic states of our model. We
shall see that pure states are in fact not pure at all. They are
statistical mixtures of Gaussian fluctuations of ontic
(``individual'')  states (prequantum classical fields). We just
remind that many authors, see, e.g. L. Ballentine [28], define the
quantum model in the same way as we did, i.e.,  without considering
pure quantum states.

\medskip

\subsection{Postulates of classical $\to$ quantum correspondence: review}
Readers who are not interested in various ``NO-GO'' theorems and who
are ready to accept the possibility to construct a prequantum
classical statistical model can omit this section. Those who still
have doubts in such a possibility (under influence of, e.g., von
Neumann or Bell ``NO-GO'' theorem) can start with our detailed
analysis of fundamental assumptions of main `NO-GO'' theorems that
is presented in this section.

As was already pointed out, we are looking for a classical
statistical model $M=(S(\Omega), V(\Omega))$ inducing the quantum
statistical model $N_{\rm{quant}} =({\cal D}(H_c), \ls(H_c)).$ The
main problem is that the meaning of the term ``inducing'' was not
specified!  For example, one may postulate (see e.g. von Neumann
[4], p. 313) that

\medskip

{\bf Postulate VO} {\it There is one to one correspondence between the space of variables $V(\Omega)$ and the space of observables $\ls(H_c).$}

\medskip

In such a case one could define a one-to-one map:
\begin{equation}
\label{M3}
T: V(\Omega) \to \ls(H_c)
\end{equation}

One can also postulate that (see e.g. von Neumann [4] p. 301-305):

{\bf Postulate SS} {\it Each quantum statistical state $D \in {\cal
D}$ corresponds to a classical statistical state $\rho \in S$ which
is uniquely defined.}

Thus there is given a map
\begin{equation}
\label{M300}
T: S(\Omega) \to {\cal D}(H_c)
\end{equation}
Moreover,  there is often postulated (see e.g. Theorem of  Kochen and Specker [14]; in von Neumann book [4] it can be derived from equality $({\bf Dis_2})$, p. 313):

{\bf Postulate F} {\it Let $g: {\bf R} \to {\bf R}$ be a Borel
function such that, for any variable $f\in V, g(f)\in V.$ Then
$T(g(f)) = g(T(f)).$}

Both models under consideration -- a classical model (for which we are looking for) and the quantum model $N_{\rm{quant}}$ --
are statistical; the final outputs of both models are averages:
$< f >_\rho$ and $<A>_D,$ which are defined by (\ref{AV0}) and (\ref{AV1}), respectively. One could postulate
(see e.g. von Neumann [4], p.301) that:

\medskip

{\bf Postulate AVC} {\it Classical and quantum averages  coincide}

\medskip

In such a case one has:
\begin{equation}
\label{AV3}
< f >_\rho = <A>_D, \; A=T(f), D= T(\rho).
\end{equation}  Thus
\begin{equation}
\label{AV4} \int_\Omega f(\psi) d \rho(\psi)=  \Tr \; D A, \; \;
A=T(f), D= T(\rho).
\end{equation}

As was mentioned, these postulates were considered, in particular, by J. von Neumann. Finally, he also postulated that

\medskip

{\bf Postulate AD} {\it The correspondence map $T$ is additive:
\begin{equation}
\label{AD}
T(f_1+ ...+ f_n+ ...) = T(f_1)+ ...+  T(f_n)+ ...,
\end{equation}
for any sequence of variables $f_1,..., f_n ,... \in  V(\Omega).$}
\footnote{It is important to remark that J. von Neumann did not assume that observables
$T(f_1), ...,  T(f_n), ...$ could be measured simultaneously!}

\medskip

Already in 30th J. von Neumann demonstrated that a correspondence map $T$ satisfying to Postulates {\bf VO, SS, F, AVC, AD} does not exist [4]. J. Bell [12] paid attention to the fact that not all
von Neumann's postulates were physically justified. He (and not only he, see L. Ballentine [15], [29] for details) strongly  criticized  Postulate {\bf AD} as totally nonphysical [12]. J. Bell also strongly criticized  the
Postulate {\bf VO}. He pointed out that it might happen that a few different physical variables are mapped
into the same physical observable. He proposed to eliminate Postulates {\bf VO, AD}  and even consider, instead of the Postulate {\bf F},  a weaker condition:

\medskip

{\bf Postulate RVC} {\it  Ranges of values of a  variable
$f \in V$ and the corresponding  quantum observable $A=T(f)$ coincide.}

\medskip

Then he proved [12] that there is still no such a correspondence map $T.$  Nevertheless, let us suppose that a prequantum
ontic model exists. It is natural to ask following questions:

\medskip

``Which postulate does block the construction of the correspondence map $T?$
Which postulate is really nonphysical?''

\medskip

\subsection{On correspondence between ranges of values of classical variables
and quantum observables} We emphasize that physical variables $f \in
V$ and observables $F \in O$ are defined on different sets of
parameters and  therefore they could have different ranges of
values, see H. Stapp [58] for detailed analysis of this problem. In
general, a measurement process induces some loss of information
about the (ontic) state $\psi.$\footnote{In fact, quantum
measurements induces huge loss of information in process of
extracting information about properties of microscopic structures
with the aid of macroscopic measurement devices.} Therefore an
observable is only an approximation of a physical variable. It seems
that the Postulate {\bf RVC}  is nonphysical (and consequently its
stronger form --  Postulate {\bf F}). By rejecting these postulates
we escape in particular the problem with the violation of Bell's
inequality.

It is also very important to remark that our measurement devices
works as {\it amplifiers} of micro effects. To be visible (e.g. at
the pointer of an apparatus) micro effects should proceed through
huge (practically infinite) amplifications. Of course, such
amplifications would totally change the ranges of values of
classical variables. These amplifications are taken into account in
our model, see (\ref{AMP}). In our approach QM is not about
microworld as it is, but about results of our measurements on it.
This is precisely the viewpoint of N. Bohr, W. Heisenberg, W. Pauli
(so called Copenhagen interpretation). The crucial difference from
the Copenhagen interpretation is that in our approach QM could be
completed by a deterministic theory, describing the motion of
individual systems. But these are  systems with the infinite number
of degrees of freedom -- ``classical fields.''

\subsection{Asymptotic equality of classical and quantum averages} The crucial
point is that in our approach, instead of the equality (\ref{AV3}),
we have the following asymptotic equality of classical and quantum
averages:
 \begin{equation}
\label{AQ} < f >_\rho = \alpha <T(f)>_{T(\rho)} + o(\alpha), \; \;
\alpha \to 0
\end{equation}
(here $<T(f)>_{T(\rho)}$ is the  quantum average).
 In mathematical models this equality has the form:
\begin{equation}
\label{AQ1} \int_\Omega f(\psi) d \rho(\psi)=  \alpha \; \Tr \; D A
+ o(\alpha), \; \; A=T(f), D= T(\rho).
\end{equation}
This equality can be interpreted in the following way. Let $f(\psi)$
be a classical physical variable (describing properties of
microsystems - classical fields having very small magnitude
$\alpha).$  We define its {\it amplification} by:
 \begin{equation}
\label{AMP} f_\alpha (\psi) =\frac{1}{\alpha} f(\psi)
\end{equation}
(so any micro effect is amplified in $\frac{1}{\alpha}$-times). Then
we have:
\begin{equation} \label{AQ4} < f_\alpha >_\rho =
<T(f)>_{T(\rho)} + o(1), \; \; \alpha \to 0,
\end{equation}
or
\begin{equation} \label{AQ5}
\int_\Omega f_\alpha(\psi) d \rho(\psi)=   \Tr \; D A + o(1), \; \;
A=T(f), D= T(\rho).
\end{equation}
Thus: {\it Quantum average $\approx$ Classical average of the
$\frac{1}{\alpha}$-amplification.} Hence: {\it QM is a mathematical
formalism describing a statistical approximation of amplification of
micro effects.}

To distinguish statistical and dynamical problems, in this paper we
shall consider the case of the real Hilbert space $H.$ Thus in all
above considerations the complex Hilbert space $H_c$ should be
changed to the real Hilbert space $H.$ In particular, $\ls\equiv
\ls(H), {\it D}\equiv {\cal D}(H).$ The case of the complex Hilbert
state space will be considered in the next paper. We start with very
short presentations of some mathematical structures on
infinite-dimensional spaces.

\section{Gaussian measures on Hilbert spaces}

Let $H$ be a real Hilbert space and let $A: H \to H$ be a continuous self-adjoint linear operator.
The basic mathematical formula which will be used in this paper is the formula for a Gaussian integral of a quadratic form
\begin{equation}
\label{QF} f(\psi)\equiv f_A (\psi)= (A\psi, \psi).
\end{equation}
Let $d\rho(\psi)$ be a $\sigma$-additive Gaussian measure on the
$\sigma$-field $F$ of Borel subsets of $H.$ This measure is
determined by its covariation operator $B: H\to H$ and mean value
$m\equiv m_\rho \in H.$ For example, $B$ and $m$ determines the
Fourier transform of $\rho:$
$$
\tilde \rho (y)= \int_H e^{i(y, \psi)} d\rho (\psi)=
e^{\frac{1}{2}(By, y) + i(m, y)}, y \in H.
$$
In what follows we restrict our considerations to Gaussian measures with zero mean value $m=0,$ where
$$
(m,y) = \int_H (y, \psi) d\rho (\psi)= 0
$$
for any $y \in H.$ Sometimes there will be used the symbol $\rho_B$ to denote the Gaussian measure with the covariation operator $B$ and $m=0.$ We recall that the covariation operator $B\equiv \rm{cov} \; \rho$ is defined by
\begin{equation}
\label{CO} (By_1, y_2)=\int (y_1, \psi) (y_2, \psi) d\rho(\psi),
y_1, y_2 \in H,
\end{equation}

and has the following properties:

a). $B \geq 0,$ i.e., ($By, y) \geq 0, y \in H;$

b). $B$ is a self-adjoint operator, $B \in \ls(H);$

c). $B$ is a trace-class operator and
\begin{equation}
\label{TR}{\rm Tr}\; B=\int_H ||\psi||^2 d\rho(\psi)
\end{equation}

The right-hand side of (\ref{TR}) defines {\it dispersion}  of the probability $\rho.$ Thus for a Gaussian
probability we have
\begin{equation}
\label{TR0}
\sigma^2(\rho)= {\rm Tr}\; B.
\end{equation}

We pay attention that the list of properties of the covariation operator of
a Gaussian measure differs from the list of properties of a von Neumann density
operator only by one condition: $\rm{Tr} \; D =1,$ for a density operator $D.$

By using (\ref{CO}) we can easily find the Gaussian integral of the
quadratic form $f_A(\psi)$ defined by (\ref{QF}):
$$
\int_H f_A(\psi) d\rho(\psi)=\int_H (A\psi, \psi) d\rho (\psi)
$$
$$
=\sum^\infty_{i, j=1}(Ae_i, e_j) \int_H (e_i, \psi) (e_j, \psi) d
\rho(\psi)= \sum_{i, j=1}^\infty (A e_i, e_j)(Be_i, e_j),
$$
where $\{e_i\}$ is some orthonormal basis in $H.$ Thus
\begin{equation}
\label{QI} \int_H f_A(\psi) d\rho (\psi)={\rm Tr}\; BA
\end{equation}

 We have presented some facts about Gaussian measures on Hilbert space; there is
 a huge number of books where one can find detailed presentation, I would like
 to recommend the excellent short book of A. V. Skorohod [59], see
 also [60]-[63] for applications to mathematical physics.

\section{Differentiable and analytic functions}

The differential calculus for maps $f: H\to {\bf R}$ does not differ
so much from the differential calculus in the finite dimensional
case, $f: {\bf R}^n \to {\bf R}.$ Instead of the norm on ${\bf
R}^n,$ one should use the norm on $H.$ We consider so called Frechet
differentiability [63]. Here a function $f$ is differentiable if it
can be represented as
$$
f(\psi_0 + \Delta \psi)= f(\psi_0) + f^\prime(\psi_0)(\Delta \psi) +
o(\Delta \psi), \; \mbox{where}\; \lim_{\Vert \Delta \psi \Vert\to
0} \frac{\Vert o(\Delta \psi)\Vert }{\Vert \Delta \psi \Vert} =0.
$$
Here at each point $\psi$ the derivative $f^\prime(\psi)$ is a
continuous linear functional on $H;$ so it can be identified with
the element $f^\prime(\psi)\in H.$ Then we can define the second
derivative as the derivative of the map $\psi\to f^\prime(\psi)$ and
so on. A map $f$ is differentiable $n$-times iff (see e.g. [63]):
$$
f(\psi_0 + \Delta \psi)= f(\psi_0) +  f^\prime(\psi_0)(\Delta \psi)
+ \frac{1}{2}f^{\prime \prime}(\psi_0)(\Delta \psi, \Delta \psi) +
...
$$
\begin{equation}
\label{Q4} +\frac{1}{n!} f^{(n)}(\psi_0)(\Delta \psi, ..., \Delta
\psi)+ o_n(\Delta \psi),
\end{equation}
where $f^{(n)}(\psi_0)$ is a symmetric continuous $n$-linear form on
$H$ and
$$
\lim_{\Vert \Delta \psi \Vert\to 0} \frac{\Vert o_n(\Delta
\psi)\Vert }{\Vert \Delta \psi \Vert^n} =0.
$$
For us it is important that $f^{\prime\prime}(\psi_0)$ can be
represented by a symmetric operator
$$
f^{\prime \prime}(\psi_0)(u,v)=(f^{\prime \prime}(\psi_0) u, v), u,
v \in H
$$
(this fact is well know in the finite dimensional case: the matrice
representing the second derivative of any two times differentiable
function $f: {\bf R}^n \to  {\bf R}$ is symmetric).
 We remark that in this case
\begin{equation}
\label{Q5} f(\psi)= f(0) +  f^\prime(0)(\psi) + \frac{1}{2}f^{\prime
\prime}(0)(\psi, \psi) + ...+ \frac{1}{n!} f^{(n)}(0)(\psi,
...,\psi) + o_n(\psi)
\end{equation}

\medskip

We recall that a functions $f: H \to {\bf R}$ is (real) analytic if
it can be expanded into series:
\begin{equation}
\label{ANN} f(\psi)= f(0) +  f^\prime(0)(\psi) +
\frac{1}{2}f^{\prime \prime}(0)(\psi, \psi) + ...+ \frac{1}{n!}
f^{(n)}(0)(\psi, ...,\psi) +... .
\end{equation}
which converges uniformly on any ball of $H,$ see [65] for details.

\section{Quantum mechanics as a projection of a classical model with infinite-dimensional state space}

Let us consider a classical statistical model in that the state
space $\Omega= H$ (in physical applications $H=L_2({\bf R}^3)$ is
the space of classical fields on  ${\bf R}^3)$ and the space of
statistical states consists of Gaussian measures with zero mean
value and dispersion
\begin{equation}
\label{DS}\sigma^2 (\rho)= \int_H \Vert \psi \Vert^2 d \rho(\psi)=
\alpha,
\end{equation}where $ \alpha> 0$ is a small real parameter. Denote such a class of Gaussian measures by the
symbol $S_G^ \alpha(H).$ For $\rho \in S_G^ \alpha(H),$ we have
\begin{equation}
\label{LL1} \rm{Tr} \; \rm{cov} \; \rho =  \alpha
\end{equation}
We remark that any linear transformation (in particular, scaling)
preserves the class of Gaussian measures. Let us make the change of
variables (scaling):
\begin{equation}
\label{LHT}  \psi \to \frac{ \psi}{\sqrt{\alpha}} .
\end{equation}
(we emphasize that this is a scaling not in the physical space ${\bf
R}^3,$ but in the space of fields on it). To find the covariation
operator $D$ of the image  $\rho_D$  of the Gaussian measure
$\rho_B,$ we compute its Fourier transform:
\[\tilde\rho_D (\xi)=\int_H e^{i(\xi, y)} d\rho_D(y)= \int_H e^{i(\xi, \frac{\psi}{\sqrt{\alpha}})} d\rho_B (\psi)=
e^{-\frac{1}{2\alpha}(B \xi, \xi)}.\] Thus
\begin{equation}
\label{LL2} D=\frac{B}{\alpha}=\frac{{\rm cov} \rho}{\alpha}.
\end{equation}
We shall use this formula later. We remark that by definition:
$$
<f>_{\rho_B} = \int_H f( \psi) d\rho_B ( \psi)= \int_H
f(\sqrt{\alpha}  \psi) d\rho_D (\psi).
$$

 Let us consider a functional space
${\cal V}(H)$ which  consists of analytic functions of  exponential growth preserving the state of vacuum:
$$
f(0)=0 \; \mbox{and there exist}\;  C_0, C_1 \geq 0 :  \vert f(
\psi)\vert \leq C_0 e^{C_1 \Vert  \psi \Vert}.
$$
We remark that any function $f\in {\cal V}(H)$ is integrable with
respect to any Gaussian measure on $H,$ see e.g. [59], [60]. Let us
consider the family of the classical statistical models
$$
M^\alpha=(S_G^\alpha(H), {\cal V}(H)).
$$
Let a variable $f \in {\cal V}(H)$ and let a statistical state
$\rho_B\in S_G^\alpha(H).$ Let us find the asymptotic expansion of
the (classical) average $<f>_{\rho_B}=\int_H f( \psi) d\rho_B(
\psi)$ with respect to the small parameter $\alpha.$ In this
Gaussian integral we make the scaling (\ref{LHT}):
\begin{equation}
\label{ANN1} <f>_{\rho_B}= \int_H f(\sqrt{\alpha}  \psi) d\rho_D (
\psi)=\sum_{n=2}^\infty  \frac{\alpha^{n/2}}{n!} \int_H f^{(n)}(0)(
\psi, ..., \psi)d\rho_D (\psi), \end{equation}where the covariation
operator $D$ is given by (\ref{LL2}). We remark that
$$
\int_H (f^\prime(0),  \psi) d\rho( \psi)=0,
$$
because the mean value of $\rho$ is equal to zero. Since $\rho_B\in
S_G^\alpha(H),$ we have
\begin{equation}
\label{LL19} \rm{Tr} \; D  = 1.
\end{equation}
The change of variables in (\ref{ANN1}) can be considered as scaling
of the magnitude of statistical  (Gaussian) fluctuations. Negligibly
small random  fluctuations
\begin{equation}
\label{DS2} \sigma (\rho)= \sqrt{\alpha},
\end{equation}
(where $\alpha$ is a small parameter) are considered in the new
scale as standard normal fluctuations. If we use the language of
probability theory and consider a Gaussian random variables
$\xi(\lambda),$ then the transformation (\ref{LHT}) is nothing else
than the standard normalization of this random variable (which is
used, for example, in the central limit theorem [65]):
\begin{equation}
\label{RVS} \eta(\lambda)= \frac{\xi(\lambda) - E
\xi}{\sqrt{E(\xi(\lambda) - E \xi)^2}}
\end{equation}
(in our case $E \xi=0).$ By (\ref{ANN1}) we have:
\begin{equation}
\label{ANN2} <f>_\rho=  \frac{\alpha}{2} \int_H (f^{\prime
\prime}(0)y, y) \; d\rho_D(y) + o(\alpha), \; \alpha \to 0,
\end{equation}
or
\begin{equation}
\label{ANN3} <f>_\rho =  \frac{\alpha}{2} \; \rm{Tr}\; D \;
f^{\prime \prime}(0) + o(\alpha), \; \alpha \to 0.
\end{equation}
We see that the classical average (computed in the model
$M^\alpha=(S_G^\alpha(H),{\cal V}(H))$ by using the
measure-theoretic approach) is coupled through (\ref{ANN3}) to the
quantum average (computed in the model $N_{\rm{quant}} =({\cal
D}(H),$ $\ls(H))$ by the von Neumann trace-formula).

The equality (\ref{ANN3}) can be used as the motivation for defining
the following classical $\to$ quantum map $T$ from the classical
statistical model $M^\alpha=(S_G^\alpha, {\cal V})$ onto the quantum
statistical model $N_{\rm{quant}}=({\cal D}, \ls):$
\begin{equation}
\label{Q20} T: S_G^\alpha(H) \to {\cal D}(H),
D=T(\rho)=\frac{\rm{cov} \; \rho}{\alpha}
\end{equation}
(the Gaussian measure $\rho$ is represented by the density matrix
$D$ which is equal to the covariation operator of this measure
normalized by  $\alpha$);
\begin{equation}
\label{Q30}T: {\cal V}(H) \to \ls(H), A_{\rm quant}= T(f)=
\frac{1}{2} f^{\prime\prime}(0).
\end{equation}
Our previous considerations can be presented as

\medskip

{\bf Theorem 6.1.} {\it The map $T$ defined by
(\ref{Q20}),(\ref{Q30}) is one-to-one on the space of statistical
states $S_G^\alpha(H)$ (so the von Neumann postulate {\bf SS}
holds); the map $T: {\cal V}(H) \to \ls(H)$ is linear (so the von
Neumann postulate {\bf AD} holds true) and the classical and quantum
averages are coupled by the asymptotic equality (\ref{ANN3}).}

\medskip

We emphasize that the correspondence between physical variables $f
\in {\cal V}(H)$ and physical observables $A \in \ls(H)$ is not
one-to-one.\footnote{ A large class of physical variables is mapped
into one physical observable. We can say that the quantum
observational model $N_{\rm{quant}}$ does not distinguish physical
variables of the classical statistical model $M^\alpha, $ The space
$ {\cal V}(H)$ is split into equivalence classes of physical
variables: $f \sim \;g \leftrightarrow f^{\prime\prime}
(0)=g^{\prime\prime}(0).$ Each equivalence class $W$ is
characterized by a continuous self-adjoint operator $A_{\rm quant}=
\frac{1}{2} f^{\prime\prime}(0),$ where $f$ is a representative of
physical variables from the class $W.$ The restriction of the map
$T$ on the space of quadratic observables $V_{\rm{quad}}(H)$ is
on-to-one. Of course, the set of variables ${\cal V}(H)$ can be
essentially extended (in particular, we can consider smooth
functions on the Hilbert space, instead of analytic functions).
However, we emphasize that such an extension would have no effect to
the quantum observational model.} Thus the von Neumann postulate
{\bf VO} is violated (as well as the postulates {\bf F}, {\bf RVC}
and {\bf AVC}).

{\bf Example 6.1.} Let $f_1( \psi)=(A  \psi,  \psi)$ and $f_2(
\psi)=\sin (A \psi,  \psi),$ where $A \in {\cal L}_s (H).$ Both
these functions belong to the space of variables ${\cal V} (H).$ In
the classical statistical model these variables have different
averages:
$$
\int_H (A \psi,  \psi) d\rho ( \psi) \not = \int_H \sin (A \psi,
\psi) d \rho ( \psi).
$$
But
$$
\int_H [(A \psi,  \psi) - \sin(A \psi,  \psi)] d\rho(x)=o(\alpha),
\alpha \to 0.
$$
Therefore by using QM we cannot distinguish these classical physical
variables. Moreover, nontrivial classical observables can disappear
without any trace in the process of transition from the prequantum
classical statistical model to QM. For example, let $f( \psi)=\cos(A
\psi,  \psi)-1.$ This is nontrivial function on $H.$ But, for any
$\rho \in S_G^\alpha(H), $ we have $<f>_\rho=o(\alpha), \alpha \to
0.$ Thus in quantum theory $f$ is identified with $g \equiv 0.$

{\bf Physical conclusions.} Our approach is based on considering the
dispersion $\alpha$ of fluctuations of classical fields  as a small
parameter. Let us consider some classical statistical model $M= (S,
V)$ and an observational model $N= (D, O).$ Suppose that this
observational model is applied to the classical one in the following
way. For any classical physical variable $f(\psi),$  there is
produced its amplification
$$
f_\alpha(\psi)= \frac{1}{\alpha}
f(\psi).
$$
Then the quantum average is defined as
\begin{equation}
\label{Q27} <A>_{\rm{quantum}}=  \lim_{\alpha\to 0}
<f_\alpha>_{\rm{classical}},
\end{equation}
where $A= \frac{1}{2} f^{\prime\prime}(0).$ So QM is a statistical
approximation of an amplification of  the classical field model (for
very small fluctuations of vacuum).

\subsection{Nonasymptotic approach}

We consider a classical statistical model  such that the class of
statistical states consists of Gaussian measures $\rho$ on $H$
having zero mean value and unit dispersion
$$
\sigma^2(\rho)= \int_H ||x|\vert^2 d\rho (x)=1.
$$
These are Gaussian measures having covariance operators with the
unit trace. Denote the class of such probabilities by the symbol
$S_G \equiv S_G(H).$ In this  model we choose a class of physical
variables consisting of quadratic forms $f_A(x)=(Ax,x).$ We denote
this class by $V_{\rm{quad}}.$ We remark that this is a linear space
(over ${\bf R}).$ We consider the following classical statistical
model:
$$
M_{\rm{quad}} = (S_G(H), V_{\rm{quad}}(H))
$$
As always in a statistical model, we are interested only in averages
of physical variables $f \in V_{\rm{quad}}$ with respect to
statistical states $\rho \in S_G(H).$ We emphasize that by
(\ref{QI}):
\begin{equation}
\label{QI0} < f_A >_\rho={\rm Tr}\; BA
\end{equation}

Let us consider the following map $T$ from the classical statistical
model $M_{\rm{quad}}= ( S_G, V_{\rm{quad}})$ to the quantum
statistical model $N_{\rm{quant}}=({\cal D}, \ls):$
\begin{equation}
\label{Q2} T: S_G(H) \to {\cal D}(H), T(\rho)= \rm{cov} \; \rho
\end{equation}
(the Gaussian measure $\rho_B$ is represented by the density matrix
$D$ which is equal to the covariation operator of this measure), and
we define:
\begin{equation}
\label{Q3}T: V_{\rm{quad}}(H) \to \ls(H), T(f)= \frac{1}{2}
f^{\prime\prime}(0)
\end{equation}
(thus a variable $f\in V_{\rm{quad}}(H)$ is represented by its
second derivative).

{\bf Theorem 6.1a.} {\it The map $T$ provides one-to-one
correspondence between the classical statistical model
$M_{\rm{quad}}$ and the quantum model $N_{\rm{quant}}.$}

Here the von Neumann postulate {\bf VO},  {\bf AD} and {\bf SS} hold
true, but the postulates {\bf F}, {\bf RVC}, {\bf AVC} are violated.
 Such a solution of the problem of classical$\to$ quantum
correspondence is acceptable mathematically. But from the physical
viewpoint  the asymptotic approach is more adequate to reality,
because quantum effects are produced by small fluctuations around
the vacuum field.

\section{Discussion}

\subsection{The statistical origin of Gaussian prequantum states}
The choice of Gaussian probability distributions as statistical
states is natural from the probabilistic viewpoint. By the central
limit theorem (which is also valid for $H$-valued random variables,
see [65]) a Gaussian probability distribution appears as the
integral effect of infinitely many independent random influences. Of
course, it is important that in our case each random influence is
given by a random variable $\xi(\psi) \in H.$ Thus we consider the
infinite number of degrees of freedom. A Gaussian distribution
$\rho$ is the integral result of influences of infinitely many such
$\xi.$ But from the purely measure-theoretical viewpoint there is no
so much difference between the origin of Gaussian probability
distribution on $H$ and $\br^n$.

\subsection{Fluctuations of vacuum} The approach based on scaling of
statistical states has some interesting physical consequences. The
space $S_G^\alpha(H)$ of statistical states of the prequantum
classical model consists of Gaussian distributions with zero mean
value and dispersion of the magnitude $\alpha.$  If $\alpha$ is very
small, then such a $\rho$ is concentrated in a very small
neighborhood of the field $\psi_{\rm{vacuum}}\equiv 0.$ Let us
interpret it as the {\it  vacuum field.}\footnote{We remark that
this is the classical vacuum field and not a vacuum state of QFT.}
Thus von Neumann density matrices represent Gaussian statistical
states (on the infinite dimensional state space $H$) which are very
narrow concentrated around the vacuum field. Such states can be
considered as {\it fluctuations of vacuum,} cf. [51], [52], [18].

\subsection{Applications outside the quantum domain}
 The statistical viewpoint to the small parameter $\alpha$ gives
the possibility to apply the quantum formalism in any statistical
model (in any domain of science) which contains statistical states
having dispersion of the magnitude $\alpha,$ where $\alpha$ is some
small parameter. It is clear that such a model describes very fine
effects. In a coarser approximation such statistical states would be
considered as states with zero dispersion.

\subsection{Second quantization}
Finally, we emphasize again that in fact there are two classical
statistical models: ordinary classical statistical mechanics (CSM)
on the phase space $\br^3\times \br^3$ and classical statistical
mechanics on the infinite dimensional Hilbert space. It is well
known that the latter classical mechanics can be quantized again.
This is the procedure of second quantization. This procedure gives
nothing else than operator quantization approach to QFT, see e.g.
[66]. There can also be established the principle of correspondence
between classical and quantum models for systems with the infinite
number of degrees of freedom. The easiest way do to this is to
repeat Weyl's considerations  and use the calculus of
infinite-dimensional pseudo-differential operators (PDO). Such a
calculus was developed on the physical level of rigorousness  in
[66] and on on the mathematical level of rigorousness  by O. G.
Smolyanov and the author [67], [64]; finally there was proved the
principle of correspondence, [68]. But in this paper we are not
interested in QFT. We only remark that methods developed in this
paper can be generalized to QFT which can also be presented as the
$T$-projection of a classical statistical model.

\section{Gaussian measures inducing pure quantum states: statistical meaning of the wave function}

\subsection{Gaussian underground} In QM  a pure quantum state is
given by a normalized vector $\psi \in H: \ \Vert \psi \Vert=1.$ The
corresponding statistical state is represented by the density
operator:
\begin{equation}
\label{DMP}D_\psi=\psi \otimes \psi.
\end{equation}
In particular, the von Neumann's trace-formula for expectation has the form:
\begin{equation}
\label{DMP1}
\rm{Tr}\;  D_\psi A=(A \psi, \psi).
\end{equation}
Let us consider the correspondence map $T$ for statistical states
for the classical statistical model $M^\alpha=(S_G^\alpha, {\cal
V}),$ see (\ref{Q20}). A pure quantum state $\psi$ (i.e., the state
with the density operator $D_\psi)$ is the image of the Gaussian
statistical mixture $\rho_\psi$ of states $\phi \in H.$ Here the
measure $\rho_\psi$ has the covariation operator
\begin{equation}
\label{DMP0} B_\psi= \alpha D_\psi.
\end{equation}
Thus
$$
(B_\psi y_1, y_2)= \int_H (y_1, \phi) (y_2,  \phi) d \rho \psi (
\phi) = \alpha (y_1, \psi) (\psi, y_2).
$$
This implies that the Fourier transform of the measure $\rho_\psi$ has the form:
$$
\tilde \rho_\psi (y)=e^{-\frac{ \alpha}{2}(y, \psi)^2}, y \in H.
$$
This means that the measure $\rho_\psi$ is concentrated on the one-dimensional subspace
$$
H_\psi=\{x \in H: x=s\psi, s \in \br\}.
$$
This is one-dimensional Gaussian distribution. It is very important to pay attention to the trivial
mathematical fact:

\medskip

{\it Concentration on the one-dimensional subspace $H_\psi$ does not imply that the Gaussian measure $\rho_\psi$ is a
pure state of type of the Dirac $\delta$-function on the classical state space $\Omega= H.$}

\medskip

\subsection{Ontic states and wave functions} In our ontic model
states are represented by vectors of the Hilbert space $H.$ Since
pure states in QM are also represented by vectors of $H,$ one might
try to identify them. The important difference is that any vector
belonging to $H$ represents an ontic state, but only normalized
vectors of $H$ represent pure quantum states. However, this is not
the crucial point. The crucial point is that the von Neumann density
operator $D_\psi=\psi \otimes \psi$ has nothing to do with the ontic
state $\psi,$ even in the case of $\Vert \psi\Vert =1.$ The density
operator describes not an individual state, but a Gaussian
statistical ensemble of individual states. States in this ensemble
can have (with corresponding probabilities) any magnitude.

\medskip

{\it Quantum pure states $\psi \in H, ||\psi||=1, $ represent
Gaussian statistical mixtures of classical states $\phi\in H.$
Therefore, quantum randomness is ordinary Gaussian randomness (so it
is reducible to the classical ensemble randomness).}

\section{Incompleteness}

Assume that our classical statistical statistical model provides the
adequate description of physical reality. This would imply that {\it
quantum mechanics is not complete} -- since it does not describe
``individual states'' $\psi \in \Omega.$ However, it seems that it
is practically impossible to verify this prediction experimentally,
because it is impossible to prepare ``pure ontic states'' $\psi$ for
microscopic systems. It is  easier to prove that quantum mechanics
is not complete even as a statistical model, namely that in nature
there exist classical statistical states (different from
$\delta_\psi$-states) which have no image in quantum model.

Let us start with ``pure nonquantum states.'' Let $\psi \in H,$ but its norm need not be equal to one. Let us consider the corresponding Gaussian statistical state $\rho_{\psi},$ see (\ref{DMP0}). This state represents the Gaussian distribution concentrated on the real line. We pay attention that by scaling  the vector $\psi$ we obtain a completely different Gaussian distribution. The only commonality between measures $\rho_{\psi}$ and $\rho_{\lambda \psi}, \lambda \in {\bf R},$
is that they are concentrated on the same real line. But they have different dispersions (and so shapes).
In particular, it is impossible to represent all scalings by the normalized vector $\psi/\Vert \psi \Vert.$

Suppose now that $\Vert \psi \Vert =o(1),  \alpha \to 0.$ In our
mathematical model there exist classical statistical states
$\rho_\psi$ with covariance matrices $B_\psi=  \alpha \psi \otimes
\psi,$ see (\ref{DMP0}). However, the quantum statistical model
$N_{\rm{quant}}$ does not contain images of such states, because
$\sigma^2(\rho)= o( \alpha).$

In the same way we can consider any classical statistical state
$\rho$ having the dispersion $\sigma^2(\rho)$ such that:
$\sigma^2(\rho)= o( \alpha).$

\section{Interpretation and comparing with other realistic prequantum models}

\subsection{Ensemble interpretation.}  In our model, PCSFT,
basic elements of  reality  are systems with the infinite number of
degrees of freedom, say classical fields. Thus our model is
 statistical mechanics of classical fields.
 Statistical states which
that correspond to statistical states described by quantum mechanics
are Gaussian distributions of such fields.   The mean value of these
Gaussian fluctuations is the vacuum field, $\psi_{\rm{vacuum}}
\equiv 0.$

We use the ensemble (or statistical) interpretation of quantum
states, since they are images, $D=T(\rho),$ of Gaussian statistical
states. The only difference from the conventional ensemble (or
statistical) interpretation of quantum mechanics,  cf. Einstein,
Margenau, Ballentine [15], [28] is that we consider ensembles of
classical fields, instead of ensembles of particles. We pay
attention that, as well as in the statistical interpretation of
Einstein, Margenau, Ballentine, even so called pure states of
quantum mechanics represent states of  ensembles of systems (in our
case classical fields) and not states of individual systems.

\subsection{Comparing with  views of Schr\"odinger.} Our views are close to
views of Schr\"odinger's original views to the wave function as a
classical scalar field as well as his later ideas to exclude totally
particles from quantum mechanics, see [6], [7]. However, the latter
program was performed in QFT-framework. In  PCSFT we do not consider
quantized fields. So it seems that the problem of even simpler that
it was considered by Schr\"odinger: we need on appeal to theory of
quantized fields it is enough to consider statistical mechanics for
classical fields (so this is really the viewpoint of Schr\"odinger
in 1920th).

\subsection{Comparing  PCSFT with Bohmian mechanics.} The main difference
between  PCSFT and the Bohmian model is that the Bohmian model still
contains particles as fundamental objects. In particular, quantum
randomness is due to randomness of initial states of particles and
not randomness of initial states of fields as in PCSFT.
Nevertheless, the presence of a field element, namely the pilot
wave, induces some similarities between Bohmian mechanics and PCSFT.

Moreover, in Bohmian mechanics particles are well defined only for
Fermi-fields, but bosons, e.g., photons cannot be defined as
corpuscular objects, but only as fields, cf. with PCSFT.\footnote{I
am thankful to Andrei Grib for this comment on Bohmian mechanics.}

\subsection{Comparing  PCSFT with   Nelson's stochastic quantum mechanics and SED}
Here comparing is very similar to comparing with Bohmian mechanics:
particles are fundamental elements of SED and Nelson's stochastic
quantum mechanics, but not of  PCSFT. In SED and Nelson's stochastic
quantum mechanics quantum randomness is the result of interaction of
particles with random media (``fluctuations of vacuum''). In  PCSFT
particles by themselves are images of fluctuating fields. So the
crucial point is not the presence of fluctuations of vacuum which
disturb the motion of corpuscular objects, but that behind any
``quantum particle'' there is a classical (``prequantum'') field.

In  PCSFT  the field of real vacuum, $\psi_{\rm{vacuum}}\equiv 0,$
has zero energy, because for any quadratic form ${\cal H}(\psi)=
({\bf H} \psi, \psi),$ where ${\bf H}$ is a self-adjoint operator,
we have $ {\cal H}(\psi_{\rm{vacuum}})=0. $ In fact, in PCSFT an
analogue of the zero point field  is the Gaussian ensemble of fields
$\psi(x)$ described by the classical statistical state $\rho.$ In
PCSFT there are numerous ``zero point fields'' $\rho$ which
reproduce quantum states. There is a similar  point in PCSFT and SED
and Nelson's mechanics, namely the dispersion of energy of the the
zero point field (Gaussian measure $\rho$ in PCSFT) is the basic
element of the model.

First we recall the situation in SED [51], [52]. Here the
fluctuations of this energy are taken as a fact. They occur  in QED
-- where they are given a formal treatment. SED provides detailed
analysis of the impact of these fluctuation, but SED could not
provide any independent justification. For a single mode of the
field, the dispersion of the electric field in the vacuum state is
$hf/2V,$ where $f$ is the frequency and $V$ the normalization volume
(in a discrete description). The corresponding spectral energy
density becomes therefore proportional to $f^3.$ There are strong
reasons to support this as the spectrum of the zero point field,
among others, that it is the only Lorentz-invariant solution, and
thus the only one consistent with the law of inertia. But if the
whole field, with all its frequency components (from zero to
infinity) is taken into account, the dispersion becomes obviously
{\it infinite,} with $f$ going to infinity (the same reason for the
old ultraviolet catastrophe). Typically one inserts a (somewhat
artificial) cutoff of  frequency. The cutoff is taken as meaningful
by some authors, by referring to a high-frequency limit in the
response of particles to the field fluctuations. Others take a more
pragmatic view by referring to the observed or measured
fluctuations, not to the physical ones, those existing in nature.
The crucial difference between SED and PCSFT is that in the latter
theory vacuum fluctuations have essentially less magnitude. In PCSFT
the total dispersion (i.e., integrated over all frequencies) is
finite! There is no analogue of ultraviolet divergences. The
integral
\begin{equation}
\label{I} \int_{0}^\infty d f \int_{L_2({\bf R}^3)} \vert
\psi(f)\vert^2 d\rho (\psi) = \sigma^2 (\rho) =\alpha< \infty
\end{equation}
Thus with respect to the energy-scale PCSFT is not only a prequantum
model, but it is even a pre-SED model. At the moment we do not know
the magnitude of $\alpha,$ i.e., a prequantum  energy scale.

We also pay attention that there exist (at least in the mathematical
model) statistical Gaussian states representing fluctuations of the
vacuum field with the statistical deviation $\sigma^2(\rho)=
o(\alpha).$  Such statistical states are neglected in the modern
observational model, QM, in that there are taken into account only
states with $\sigma^2(\rho)= \alpha.$ But statistical states which
we neglect  in QM have nonzero average of energy:
 $$
 <{\cal H}>_\rho= \int_H ({\bf H} \psi, \psi) d \rho (\psi).
 $$
 Of course, this average is negligibly small:
 $$
  \vert<{\cal H}>_\rho\vert\leq \Vert {\bf H}\Vert \int_H \Vert \psi\Vert^2 d \rho (\psi)=
  \Vert {\bf H}\Vert \sigma^2 (\rho)= o(\alpha)
 $$
(we considered the case of continuous operator ${\bf H}: H \to H).$
We can say that PCFT supports the zero point field model.

The main open question in comparing Nelson's stochastic mechanics
and SED with PCSFT is about the magnitude $\alpha$ of vacuum
fluctuations in PCSFT.

\section{Finite-dimensional QM as an image of CSM} Let us consider
apply our approach to the finite-dimensional case. We consider the
classical statistical model $ M^\alpha({\bf R}^n)= (S_G^\alpha({\bf
R}^n), {\cal V}({\bf R}^n)). $ This is a special sub-model of
classical statistical mechanics. The conventional model is given by
$ M = (PM({\bf R}^n)), C^{\infty}_b({\bf R}^n) ). $ Here the space
of statistical states $PM({\bf R}^n))$ coincides with the space of
all probability measures and the space of physical variables
$C^{\infty}_b({\bf R}^n)$ consists of smooth bounded functions.

We emphasize again that the choice of a special class of statistical
states is crucial to obtain a quantum-like representation. If one
chooses the class of statistical states consisting of all
probability measures, then it would be impossible to project it onto
QM.

 Let us now consider a variant of QM in
that the state space is finite-dimensional. As we consider in this
paper only real numbers, so in the present section $ H={\bf R}^n. $
We consider the quantum model
$$
N_{\rm{quant}}({\bf R}^n)= ({\cal D}({\bf R}^n), {\cal L}_s({\bf
R}^n) ).
$$
By using the Taylor expansion (now on the ${\bf R}^n)$ we can
establish the $T$-correspondence between the models $M^\alpha({\bf
R}^n)$ and $N_{\rm{quant}}({\bf R}^n)$ and obtain the fundamental
equality:
$$
<f>_\rho= \alpha \; <T(f)>_{T(\rho)} + o(\alpha), \;
\alpha\to 0.
$$

In particular, one can imagine a super-macroscopic observer such
that our macroscopic quantities are negligibly small for him. By
introducing a parameter $\alpha$ (which is  small for him, but
sufficiently large for us) he can create a representation of
classical statistical mechanics on ${\bf R}^n$ in the form of the
quantum model with a finite-dimensional state space. He can make the
$1/\alpha$-amplification of our classical physical variable
$f(\psi)$ and define his ``super-quantum'' average.

This approach need not be applied to physics. Any collection of
 statistical data on ${\bf R}^n$
(e.g., in economics) can be represented in the quantum-like way
through introducing a small parameter $\alpha$ giving us the
precision of calculating of averages.

\section{Extension of the space of statistical states}
We have seen that the quantum (observational) statistical model can
be considered as the image of a classical (ontic) statistical model.
In our classical model the space of statistical states consists of
Gaussian distributions having zero mean value and dispersion
$\sigma^2(\rho)=\alpha.$ Such states describes Gaussian fluctuations
of the  vacuum field. The statistical magnitude of fluctuations is
equal to $\alpha.$ However, in all our considerations there was
important only the magnitude of fluctuations is {\it approximately
equal $\alpha.$} Therefore we can essentially extend the class of
Gaussian classical statistical states and still obtain the same set
of quantum states ${\cal D}(H).$ Of course, for such a model the
correspondence between classical and quantum statistical states
would not be one-to-one. Let us consider the space of Gaussian
measures on H having zero mean value and dispersion.
\begin{equation}
\label{DH} \sigma^2(\rho)= \alpha + o(\alpha), \alpha \to 0.
\end{equation}
Denote it by the symbol $S_G^{\approx \alpha}(H)$. We consider the
following correspondence map between classical and quantum
statistical states extending the map (\ref{Q20}):
\begin{equation}
\label{Rome} T: S_G^{\approx \alpha}(H) \to {\cal D} (H),
T(\rho)=\frac{{\rm cov}\rho}{\sigma^2(\rho)}
\end{equation}
We see that the operator $D=T(\rho) \in {\cal D} (H),$ so the map
$T$ is well defined.

{\bf Proposition 11.2.}{\it \; For the map $T$ defined by
(\ref{Rome}) the asymptotic equality of classical and quantum
averages (\ref{ANN3}) holds for any variable $f \in {\cal V} (H).$}

{\bf Proof.} We have $<f>_{\rho_B}=\int_H f(x) d\rho_B (x)=\int_H
f(\sigma(\rho_B)y) d\rho_D(y)=\frac{\alpha+o(\alpha)}{2} \int_H
(f^{\prime\prime}(0)y, y) d\rho_D(y) + o(\alpha).$ So we obtained
the asymptotic equality (\ref{ANN3}).

\medskip

As was pointed out, two different Gaussian measures $\rho_1, \rho_2
\in S_G^{\approx \alpha}(H)$ can be mapped to the same density
operator $D.$ If the condition
\begin{equation}
\label{Rome1} \sigma^2(\rho_1) - \sigma^2(\rho_2)=o(h), h \to 0,
\end{equation}
holds, then $T(\rho_1)= T(\rho_2).$ Here even the von Neumann
postulate {\bf SS } is violated and only the postulate {\bf AD}
(which was strongly criticized by J. Bell) still holds true.

 Our choice of
Gaussian statistical states is based on central limit theorem for
$H$-valued independent random variables (independent random
fluctuations of vacuum). However, in principle, we could not exclude
the
 possibility that in nature may exist stable non-Gaussian statistical states. We recall that the formula (\ref{QI})
giving the trace-expression of integrals of quadratic forms is valid
for arbitrary measure $\mu$ on $H$ having zero mean value and the
finite second moment:
$$
\sigma^2(\mu)=\int_H||x||^2 d\mu(x)<\infty.
$$
Denote the set of such probability measures by the symbol $PM_2(H).$
Let us consider the classical statistical model
$$
M_2^\alpha=(PM_2^\alpha(H), {\cal V}_2(H)),
$$
where $PM_2^\alpha(H)$ consists of $\mu \in PM_2(H)$ having the
dispersion $\sigma^2(\mu)=\alpha$ and the space of variables ${\cal
V}_2(H)$ consists of real analytic functions $f:H \to {\bf R},
f(0)=0,$ having
 quadratic growth for $x\to \infty:$
$$
|f(x)|\leq c_1 + c_2||x||^2, x \in H, c_1, c_2 >0.
$$
We find the average of $f \in {\cal V}_2(H)$ with respect to $\mu
\in PM_2^\alpha (H):$
\\$<f>_\mu=\int_H f(x) d\mu(x)=\int_H f(\sigma (\mu)y) d \nu (y)$\\$
=\frac{\sigma^2(\mu)}{2} \int_H (f^{\prime\prime}(0)(y, y) d \nu (x)
+ \sum_{n=2}^\infty \frac{\sigma^2(\mu)}{(2n)!} \int_H f^{(2n)}(0) (y, \ldots y) d\nu (x),$\\
where a measure $\nu$ is the scaling of the measure $\mu$ induced by
the map: $y=\frac{x}{\sigma(\mu)}.$ We remark that the covariation
operator of the measure $\nu$ is obtained as the scaling of the
covariation operator of the measure $\mu: D=\rm{cov} \nu=
\frac{\rm{cov} \mu}{\sigma(\mu)}.$

Thus we again have: $<f>_\mu=\frac{\alpha}{2}\int_H
(f^{\prime\prime}(0)y, y)d\nu(x) + o(\alpha)=\frac{\alpha}{2} {\rm
Tr \; cov} \nu f^{\prime\prime}(0) + o(\alpha).$

Hence, the quantum model $N_{\rm{quant}}$ can be considered as the
image of the classical model $M_{2}^\alpha$ and classical and
quantum averages are equal asymptotically, $\alpha\to 0.$ The map
$T$ has huge degeneration on the space of statistical states, since
a covariation operator does not determine a measure uniquely.

As well as in the Gaussian case, we can consider the space of
measures which dispersion is only approximately equal $h:$
\[PM_2^{\approx \alpha}(H)=\{\mu \in PM_2 (H): \sigma^2(\mu)= \alpha + o( \alpha),  \alpha \to 0\}.\]
The map $T$ can be extended to this class
(by increasing degeneration).

\section{Generalized quantum mechanics: approximations of higher orders}

We have created the classical statistical model which induced the
quantum statistical model. The quantum  description can be obtained
through the Taylor expansion of classical physical variables up to
the terms of the second order. The crucial point is the presence of
a parameter $\alpha$ which small in QM, but not in the prequantum
classical model.

This viewpoint to conventional  quantum mechanics implies the
evident possibility to generalize this formalism by considering
higher orders of the  Taylor expansion of classical physical
variables and corresponding expansions of classical averages with
respect to the parameter $\alpha.$

We recall that momentums of a measure $\rho$ are defined by
$$
a_\rho^{(k)} (z_1, \ldots, z_k)=\int_H (z_1, x)...(z_k, x) d \rho
(x).
$$
In particular, $a_\rho^{(1)}\equiv a_\rho$ is the mean value and
$a_\rho^{(2)}$ is the covariation form.  We remark that for a
Gaussian measure $\rho, a_\rho=0$ implies that all its momenta of
odd orders $a_\rho^{(k)}, k=2n + 1, n=0, 1, \ldots, $ are also equal
to zero.

Therefore the expansion of $<f>_\rho$ with respect to
$s=\alpha^{1/2}$ does not contain terms with $s^{2n + 1}.$ Hence
this is the expansion with respect to $\alpha^n(=s^{2n}), n=1,2,
\ldots$ We are able to create $o(\alpha^n)$-generalization of
quantum mechanics through neglecting by terms of the magnitude
$o(\alpha^n), \alpha \to 0 (n=1,2, \ldots)$ in the power expansion
of the classical average. Of course, for $n=1$ we obtain the
conventional quantum mechanics.  Let us consider the classical
statistical model
 \begin{equation}
\label{TRHJ6} M^\alpha=(S_G^\alpha (\Omega), {\cal V}(\Omega)),
\end{equation}
where $\Omega=H$ is the real Hilbert space. By taking into account
that $a_\rho^{2n + 1}=0, n=0, 1, \ldots, $ for $\rho \in
S_G^h(\Omega),$ we have:
\begin{equation}
\label{GI2} <f>_\rho=\frac{\alpha}{2} {\rm Tr} \; D f^{\prime
\prime} (0) + \sum_{k=2}^\infty \frac{\alpha^k}{(2k)!} \int_H
f^{(2k)} (0) (y, \ldots, y) d \rho_D (y),
\end{equation}
where as always $D=\frac{{\rm cov} \rho}{\alpha}.$

We now consider a new epistemic (``observational'') statistical
model which is a natural generalization of the conventional quantum
mechanics. We start with some preliminary mathematical
considerations. Let $A$ and $B$ be two $n$-linear symmetric forms.
We define their trace by
\begin{equation}
\label{QPJ6} {\rm Tr} \;  B A=\sum_{j_1, \ldots, j_n=1}^\infty
B(e_{j_1}, \ldots, e_{j_h}) A(e_{j_1}, \ldots, e_{j_n}),
\end{equation}
if this series converges and its sum does not depend on the choice
of an orthonormal basis $\{e_j\}$ in $H.$ We remark that
\begin{equation}
\label{H} <f>_\rho=\frac{\alpha}{2} {\rm Tr} \; D f^{\prime \prime}
(0) + \sum_{k=2}^n \frac{\alpha^k}{2k!} {\rm Tr} \;a_{\rho_D}^{(2k)}
f^{(2k)} (0) + o(\alpha^n), \alpha \to 0,
\end{equation}
Here we used the following  result about Gaussian integrals:

{\bf Lemma 12.1.} {\it Let $A_k$ be a continuous $k$-linear form on
$H$ and let $ \rho_D$ be a Gaussian measure (with zero mean value
and the covariation operator $D).$ Then}
\begin{equation}
\label{AKH} \int_H A_k (\psi, \ldots, \psi) d \rho_D (\psi)={\rm Tr}
\; a_{\rho_D}^{(k)} A_k.
\end{equation}

{\bf Proof.} Let $\{e_j\}_{j=1}^\infty$ be an orthonormal basis in
$H.$ We apply the well known Lebesque theorem on majorant
convergence. We set
\begin{equation}
\label{QPJ5} f_N(\psi)=\sum_{j_1, \ldots, j_k=1}^n A_k(e_{j_1},
\ldots, e_{j_k}) (e_{j_1}, \psi) \ldots (e_{j_k}, \psi).
\end{equation}
We have
\begin{equation}
\label{QPJ4}|f_N(\psi)|=|A_k (\sum_{j_1=1}^N (x, e_{j_1}) e_{j_1}
\ldots, \sum_{j_k=1}^N (\psi, e_{j_k}) e_{j_k})|\leq||A_k|| \;
||\psi||^k.
\end{equation}
Therefore we obtain:
$$
\int_H A_k (\psi, \ldots, \psi) d\rho_D(\psi)= \lim_{N\to \infty}
\int_H f_N(\psi) d\rho_D(\psi)
$$
\begin{equation}
\label{QPJ13} = \sum_{j_1=1,..., j_k=1}^\infty A_k (e_{j_1}, \ldots,
e_{j_k}) \int_H (e_{j_1}, \psi)\ldots (e_{j_k}, \psi)
d\rho_D(\psi)={\rm Tr}\; a^{(k)}_{\rho_D} A_k.
\end{equation}
The proof is finished.
\bigskip

In particular, we obtained the following inequality:
\begin{equation}
\label{QPJ2} \vert {\rm Tr} \; a^k_{\rho_D} A_k|\leq
||A||\int_H||\psi||^k d\rho_D (\psi).
\end{equation}

We now remark that for a Gaussian measure (with zero mean value)
integrals (\ref{AKH}) are equal to zero for  $k=2l+1.$   Thus ${\rm
Tr} \; a^{(2l+1)}_{\rho_D} A_{2l+1}=0.$  It is easy to see that
$2k$-linear forms (momenta of even order) $a_{\rho_D}^{2k}$ can be
expressed through the covariance operator $D:$
\begin{equation}
\label{QPJ1} a_{\rho_D}^{(2k)}=e(k,
D)=\frac{d^{2k}}{dy^{2k}}e^{-\frac{1}{2}(Dy, y)}|_{y=0}.
\end{equation}
In particular, $e(2, D)(z_1, z_2)=(Dz_1, z_2)$ and $e(4, D)(z_1,
z_2, z_3, z_4)$
\begin{equation}
\label{QPJ} = (Dz_1, z_3)(Dz_2, z_4) + (Dz_2, z_3)(Dz_1, z_4) +
(Dz_1, z_2)(Dz_3, z_4).
\end{equation}
Thus (\ref{H}) can be rewritten as
\begin{equation}
\label{EH} <f>_{\rho_B}= \frac{\alpha}{2} {\rm Tr} \;
Df^{\prime\prime}(0) + \sum_{k=2}^n\frac{\alpha^k}{2k!} {\rm Tr} \;
e(2k, D)f^{(2k)}(0) + o(\alpha^n), \; \alpha\to 0,
\end{equation}
or by introducing the $1/\alpha$-amplification of the classical
physical variable $f$ we have:
\begin{equation}
\label{EHA} <f_\alpha>_{\rho_B}=\frac{1}{2} {\rm Tr} \;
Df^{\prime\prime}(0) + \sum_{k=2}^n\frac{\alpha^{k-1}}{2k!} {\rm Tr}
\; e(2k, D)f^{(2k)}(0) + o(\alpha^{n-1})
\end{equation}

This formula is the basis of {\it a new quantum theory.} In this
theory statistical states can be still represented by von Neumann
density operators $D \in {\cal D} (H),$ but observables are
represented by
 multiples $A=(A_2, A_4, \ldots, A_{2n}),$ where $A_{2j}$
 are  symmetric $2n$-linear forms on a Hilbert space $H.$ In particular, the
 quadratic form $A_2$ can be represented by a self-adjoint operator. To escape mathematical
 difficulties, we can assume that forms $A_{2j}$ are continuous. Denote the space
 of all such multiples $A$ by $L_{2n}(H).$ We obtain the following generalization of the conventional quantum model:
\begin{equation}
\label{EHJ7} N_{\rm{quant},2n}= ({\cal D}(H), L_{2n}(H)).
\end{equation}
 Here the average of
an observable $A\in L_{2n}(H)$ with respect to a state $D\in {\cal
D}(H)$ is given by
\begin{equation}
\label{F} <a>_D=\sum_{n=1}^n {\rm Tr} \; e(2k, D) A_{2k}
\end{equation}
If one define ${\rm Tr} \; D A=\sum_{k=1}^n {\rm Tr} \; e(2k, D)
A_{2k}$ then the formula (\ref{F}) can be written as in the
conventional quantum mechanics (von Neumann's formula of $n$th
order):
\begin{equation}
\label{V} <A>_D={\rm Tr}D A
\end{equation}
This model is the result of the following ``quantization'' procedure
of the classical statistical model $M^\alpha=(S_G^\alpha (\Omega),
{\cal V}(\Omega)$:
\begin{equation}
\label{QP} \rho \to D=\frac{{\rm cov} \rho}{\alpha};
\end{equation}
\begin{equation}
\label{QP1} f\to A=(\frac{1}{2} f^{\prime\prime}(0),
\frac{\alpha}{4!}f^{(4)}(0),..., \frac{\alpha^{n-1}}{(2n)!}
f^{(2n)}(0)).
\end{equation}
(thus here $A_{2k}=\frac{\alpha^{k-1}}{(2k)!} f^{(2k)}(0)).$ The
transformation $T_{2n}$ given by (\ref{QP}), (\ref{QP1}) maps the
classical statistical model $M^\alpha=(S_G^\alpha (\Omega), {\cal
V}(\Omega))$ onto generalized quantum model $N_{\rm{quant},2n}=(H,
{\cal D} (H), L_{2n} (H)).$

In this framework one takes into account contributions to averages
up to the magnitude $\alpha^{n-1},$ but neglects by  quantities of
the magnitude $\alpha^{n}.$

\medskip

{\bf Theorem 11.1.} {\it For the classical statistical model
$M^\alpha=(S_G^\alpha(\Omega),$\\ ${\cal V}(\Omega))$, the classical
$\to$ quantum map $T_{2n},$ defined by  (\ref{QP}) and (\ref{QP1}),
is one-to-one for statistical states; it has a huge degeneration for
variables. Classical and quantum averages are coupled through the
asymptotic equality (\ref{EH})}.

\medskip

We pay attention to the simple mathematical fact that the degree of
degeneration of the map $T_{2n}: {\cal V}(\Omega)\to L_{2n}(H)$ is
decreasing for $n \to \infty.$ Denote the space of polynomials of
the degree $2n$ containing only terms of even degrees by the symbol
$P_{2n}.$ Thus $f \in P_{2n}$ iff $f(\psi)=Q_2(\psi, \psi) + Q_4
(\psi, \psi, \psi, \psi) + \ldots + Q_{2n} (\psi, \ldots ,\psi),$
where $Q_{2j}:H^{2j} \to \br$ is a symmetric $2j$-linear
(continuous) form. The restriction of the map $T_{2n}$ on the
subspace $P_{2n}$ of the space ${\cal V}$ is one-to-one. One can
also consider a generalized quantum model
\begin{equation}
\label{QPY} N_{\rm{quant},\infty}=({\cal D}, L_\infty),
\end{equation}
where $L_\infty(H)$ consists of infinite sequences of $2n$-linear
(continuous) forms on $H:$
\begin{equation}
\label{QPY1} A=(A_2, \ldots, A_{2n}, \ldots).
\end{equation}
The correspondence between the classical model $M^\alpha$ (for any
$\alpha$) and the generalized quantum model $ N_{\rm{quant},\infty}$
is one-to-one.

\bigskip

I would like to thank A. Aspect, L. Accardi, L. Ballentine, G. W.
Mackey, E. Nelsson, S. Albeverio, D. Greenberger, S. Gudder, G. `t
Hooft, Th. Nieuwenhuizen, A. Leggett, P. Lahti, A. Peres, A. S.
Holevo,
 H. Atmanspacher, K. Hess,  W. Philipp, D. Mermin,  for fruitful discussions on contextual approach to QM and its hyperbolic
generalizations.

\bigskip

{\bf References}

\bigskip

[1]  D. Hilbert, J. von Neumann, L. Nordheim, {\it Math. Ann.}, {\bf
98}, 1-30 (1927).

[2] P. A. M.  Dirac, {\it The Principles of Quantum Mechanics,}
Oxford Univ. Press, 1930.

[3]  W. Heisenberg, {\it Physical principles of quantum theory,}
Chicago Univ. Press, 1930.

[4] J. von Neumann, {\it Mathematical foundations of quantum
mechanics,} Princeton Univ. Press, Princeton, N.J., 1955.

[5] A. Einstein, B. Podolsky, N. Rosen, {\it Phys. Rev.} {\bf 47},
777--780 (1935).

[6] E. Schr\"odinger,  {\it Philosophy and the Birth of Quantum
Mechanics.} Edited by M. Bitbol, O. Darrigol (Editions Frontieres,
Gif-sur-Yvette, 1992); especially the paper of S. D'Agostino,
``Continuity and completeness in physical theory: Schr\"odinger's
return to the wave interpretation of quantum mechanics in the
1950's'', pp. 339-360.

[7]   E. Schr\"odinger, {\it E. Schr\"odinger Gesammelte
Abhandlungen} ( Wieweg and Son, Wien, 1984); especially the paper
``What is an elementary particle?'', pp. 456-463.

[8] A. Einstein, {\it The collected papers of Albert Einstein}
(Princeton Univ. Press, Princeton, 1993).

[9] A. Einstein and L. Infeld, {\it The evolution of Physics. From
early concepts to relativity and quanta} (Free Press, London, 1967).

[10]  A. Lande, {\it New foundations of quantum mechanics,}
Cambridge Univ. Press, Cambridge, 1965.

[11] L. De Broglie, {\it The current interpretation of wave
mechanics, critical study.} Elsevier Publ., Amsterdam-London-New
York, 1964.

[12] J. S. Bell, {\it Speakable and unspeakable in quantum
mechanics,} Cambridge Univ. Press, 1987.

[13] G. W. Mackey, {\it Mathematical foundations of quantum
mechanics,} W. A. Benjamin INc, New York, 1963.

[14]  S. Kochen and E. Specker, {\it J. Math. Mech.}, {\bf 17},
59-87 (1967).

[15] L. E. Ballentine, {\it Rev. Mod. Phys.}, {\bf 42}, 358--381
(1970).

[16]  G. Ludwig, {\it Foundations of quantum mechanics,} Springer,
Berlin, 1983.

[17] E. B. Davies, J. T. Lewis, {\it Comm. Math. Phys.} {\bf 17},
239-260 (1970).

[18] E. Nelson, {\it Quantum fluctuation,} Princeton Univ. Press,
Princeton, 1985.

[19] D.  Bohm  and B. Hiley, {\it The undivided universe: an
ontological interpretation of quantum mechanics,} Routledge and
Kegan Paul, London, 1993.

[20] S. P. Gudder, {\it Trans. AMS} {\bf 119}, 428-442 (1965).

[21] S. P. Gudder, {\it Axiomatic quantum mechanics and generalized
probability theory,} Academic Press, New York, 1970.

[22] S. P. Gudder, {\it ``An approach to quantum probability''} in
{\it Foundations of Probability and Physics,} edited by  A. Yu.
Khrennikov, Quantum Prob. White Noise Anal., 13,  WSP, Singapore,
2001, pp. 147-160.

[23] R. Feynman and A. Hibbs, {\it Quantum Mechanics and Path
Integrals,} McGraw-Hill, New-York, 1965.

[24] J. M. Jauch, {\it Foundations of Quantum Mechanics,}
Addison-Wesley, Reading, Mass., 1968.

[25] A. Peres, {\em Quantum Theory: Concepts and Methods,}
Dordrecht, Kluwer Academic, 1994.

[26] L. Accardi, {\it ``The probabilistic roots of the quantum
mechanical paradoxes''} in {\em The wave--particle dualism.  A
tribute to Louis de Broglie on his 90th Birthday,} edited by  S.
Diner, D. Fargue, G. Lochak and F. Selleri, D. Reidel Publ. Company,
Dordrecht, 1984, pp. 297--330.

[27] L. Accardi, {\it Urne e Camaleoni: Dialogo sulla realta, le
leggi del caso e la teoria quantistica,} Il Saggiatore, Rome, 1997.

[28]  L. E. Ballentine, {\it Quantum mechanics,} Englewood Cliffs,
New Jersey, 1989.

[29] L. E. Ballentine,  {\it ``Interpretations of probability and
quantum theory'',} in {\it Foundations of Probability and Physics,}
edited by  A. Yu. Khrennikov,
 Q. Prob. White Noise Anal.,  13,  WSP, Singapore, 2001, pp. 71-84.

[30] A. S. Holevo, {\it Probabilistic and statistical aspects of
quantum theory,} North-Holland, Amsterdam,  1982.

[31] A. S. Holevo, {\it Statistical structure of quantum theory,}
Springer, Berlin-Heidelberg, 2001.

[32] P. Busch, M. Grabowski, P. Lahti, {\it Operational Quantum
Physics,} Springer Verlag,Berlin, 1995.

[33] A. Yu. Khrennikov (editor), {\it Foundations of Probability and
Physics,} Q. Prob. White Noise Anal.,  13,  WSP, Singapore, 2001.

[34] A. Yu. Khrennikov (editor), {\it Quantum Theory:
Reconsideration of Foundations,} Ser. Math. Modeling, 2, V\"axj\"o
Univ. Press,  2002.

[35] A. Yu. Khrennikov (editor), {\it Foundations of Probability and
Physics}-2, Ser. Math. Modeling, 5, V\"axj\"o Univ. Press,  2003.

[36] A. Yu. Khrennikov (editor),  {\it Quantum Theory:
Reconsideration of Foundations}-2,  Ser. Math. Modeling, 10,
V\"axj\"o Univ. Press,  2004.

[37] H. Atmanspacher, H. Primas, {\it ``Epistemic and ontic quantum
realities''}, in {\it Foundations of Probability and Physics}-3,
edited by A. Yu. Khrennikov, AIP Conference Proceedings, 2005.

[38] A. Yu. Khrennikov, {\it Interpretations of Probability,} VSP
Int. Sc. Publishers, Utrecht/Tokyo, 1999 (second edition, 2004).

[39] A. E. Allahverdyan, R. Balian, T. M. Nieuwenhuizen, in: A. Yu.
Khrennikov (Ed.), Foundations of Probability and Physics-3,
Melville, New York: AIP Conference Proceedings, 2005, pp. 16-24.

[40] W. De Baere,  {\it Lett. Nuovo Cimento} {\bf 39}, 234 (1984);
{\bf 40}, 488(1984); {\it Advances in electronics and electron
physics} {\bf 68}, 245 (1986).

[41]  De Muynck W. M., {\it Foundations of Quantum Mechanics, an
Empiricists Approach} (Kluwer, Dordrecht) 2002.

[42] De Muynck W., De Baere W., Martens H., {\it Found. of Physics}
{\bf 24} (1994) 1589.

[43] Allahverdyan A. E., Balian R., Nieuwenhuizen Th., {\it
Europhys. Lett.} {\bf 61} (2003) 452.

[44] K. Hess and W. Philipp, {\it Proc. Nat. Acad. Sc.} {\bf 98},
14224 (2001); {\bf 98}, 14227(2001); {\bf 101}, 1799 (2004); {\it
Europhys. Lett.} {\bf 57}, 775 (2002).

[45] A. Yu. Khrennikov, {\it J. Phys.A: Math. Gen.} {\bf 34},
9965-9981 (2001); {\it Il Nuovo Cimento} {\bf B 117},  267-281
(2002); {\it J. Math. Phys.} {\bf 43}, 789-802 (2002); {\it
Information dynamics in cognitive, psychological and anomalous
phenomena,} Ser. Fundamental Theories of Physics, Kluwer, Dordreht,
2004;  {\it J. Math. Phys.} {\bf 44},  2471- 2478 (2003); {\it Phys.
Lett. A} {\bf 316}, 279-296 (2003); {\it Annalen  der Physik} {\bf
12},  575-585 (2003).

[46] A. Yu. Khrennikov, What is really "quantum" in quantum theory?
quant-ph/0301051.

[47] H. Weyl, {\it The Theory of Groups and Quantum Mechanics,}
Dover Publications, 1950.

[48]  J. E. Moyal,  {\it Proc. Camb. Phil. Soc}, {\bf 45}, 99-124
(1949).

[49] G. Dito and D. Sternheimer, Deformation quantization: genesis,
developments and metamorphoses. Deformation quantization
(Strasbourg, 2001),  9--54, IRMA Lect. Math. Theor. Phys., 1, de
Gruyter, Berlin, 2002.

[50]  A. Yu. Khrennikov,  Prequantum classical statistical model
with infinite dimensional phase-space, quant-ph/0505228, v1-v3;
Prequantum classical statistical model with infinite dimensional
phase-space-2, quant-ph/0505230.

[51] L. de la Pena and A. M. Cetto, {\it The Quantum Dice: An
Introduction to Stochastic Electrodynamics Kluwer.} Dordrecht, 1996;
T. H. Boyer, {\it A Brief Survey of Stochastic Electrodynamics} in
Foundations of Radiation Theory and Quantum Electrodynamics, edited
by A. O. Barut, Plenum, New York, 1980; T. H. Boyer, Timothy H.,
{\it Scientific American},pp 70-78, Aug 1985; see also an extended
discussion on vacuum fluctuations in: M. O. Scully, M. S. Zubairy,
{\it Quantum optics,} Cambridge University Press, Cambridge, 1997;
W. H. Louisell, {\it Quantum Statistical Properties of Radiation.}
J. Wiley, New York, 1973; L. Mandel and E. Wolf, {\it Optical
Coherence and Quantum Optics.} Cambridge University Press,
Cambridge, 1995.

[52] L. De La Pena, {\it Found. Phys.} {\bf 12}, 1017 (1982); {\it
J. Math. Phys.} {\bf 10}, 1620 (1969); L. De La Pena, A. M. Cetto,
{\it Phys. Rev. D} {\bf 3}, 795 (1971).

[53]  G. `t Hooft, ``Quantum Mechanics and Determinism,''
hep-th/0105105.

[54] G. `t Hooft,``Determinism beneath Quantum Mechanics,''
quant-ph/0212095.

[55] A. Plotnitsky, {\it The knowable and unknowable (Modern
science, nonclassical thought, and the ``two cultures)} (Univ.
Michigan Press, 2002);

[56] A. Plotnitsky, ``Quantum atomicity and quantum information:
Bohr, Heisenberg, and quantum mechanics as an information theory'',
in {\it Quantum theory: reconsideration of foundations},  A. Yu.
Khrennikov,ed.( V\"axj\"o Univ. Press,  2002), pp. 309-343.

[57] A. N. Kolmogoroff, {\it Grundbegriffe der
Wahrscheinlichkeitsrechnung} (Springer Verlag, Berlin, 1933);
reprinted: {\it Foundations of the Probability Theory} (Chelsea
Publ. Comp., New York, 1956).

[58] H. P. Stapp, Phys. Rev., D, {\bf 3}, 1303-1320 (1971).

[59] A. V. Skorohod, Integration in Hilbert space. Springer-Verlag,
Berlin, 1974.

[60] A. L. Daletski and S. V. Fomin, Measures and differential
equations in infinite-dimensional spaces. Kluwer, Dordrecht, 1991.

[61] S. Albeverio, and M. R\"ockner, {\it Prob. Theory and Related
Fields} {\bf 89}, 347 (1991).

[62] S. Albeverio, R. Hoegh-Krohn, {\it Phys. Lett.} {}bf B 177, 175
(1989).

[63]  O. G. Smolyanov, {\it Analysis on topological vector spaces}
(Moscow State Univ. Press, Moscow, 1981).

[64]  A. Yu. Khrennikov, {\it Izvestia Akademii Nauk USSR,
ser.Math.}, {\bf 51}, 46 (1987); A. Yu. Khrennikov, H. Petersson,
Ibid, {\bf 65}, N. 2, 201 (2001).

[65] V. Bentkus,   V. Paulauskas,  et al, editors, {\it Limit
Theorems of Probability Theory.} Springer-Verlag,  Berlin and
Heidelberg, 1999.

[66] F. A. Beresin, {\it The method of second quantization.} Nauka,
Moscow, 1965.

[67] O. G. Smolyanov,  {\it Dokl. Akad. Nauk USSR}, {\bf 263},
558(1982).

[68] A. Yu. Khrennikov, {\it Matematicheskii Sbornic}, {\bf 180},
763 (1989).

\end{document}